\documentclass[twocolumn]{elsart3p}

\usepackage{amssymb}
\usepackage{graphicx}
\usepackage{amsmath} 
\usepackage{color}

\newcommand{\comment}[1]{}

\def\mib#1{\mbox{\boldmath $#1$}}

\begin{document}

\begin{frontmatter}


%\author{\corauthref{cor1}\thanksref{label2}}

\title{
Cascade hypernuclear production spectra at J-PARC
%Production spectra of $\Xi^-$ hypernuclei
%by the $(K^-,K^+)$ reaction
}


\author{
Hideki Maekawa, Kohsuke Tsubakihara, Hiroshi Matsumiya and Akira Ohnishi
}

\address{
Department of Physics, Faculty of Science,
Hokkaido University Sapporo 060-0810, Japan
}

\begin{abstract}
% Text of abstract
We predict cascade hypernuclear production spectra
expected in the forthcoming J-PARC experiment.
In the Green's function method
of the distorted wave impulse wave approximation
with the local optimal Fermi averaging $t$-matrix,
we can describe the $\Xi^-$ production spectra
in the continuum and bound state region reasonably well.
Predictions
%to the high resonlution $\Xi$ production spectra at J-PARC
to the high resonlution spectra at J-PARC
suggest that we should observe $\Xi^-$ bound state peak structure
in $(K^-,K^+)$ spectra in light nuclear targets
such as $^{12}$C and $^{27}$Al.
\end{abstract}

\end{frontmatter}

% main text

\section{Introduction}
\label{intro}

Investigation of nuclear systems with strangeness
opens up a way to understand dense matter
such as the neutron star core.
Since strange quarks are negatively charged
and cancel the positive proton charge,
they are favored in charge neutral dense matter.
As a result, many models predict hyperons would appear at around $2\rho_0$,
and $\Lambda$ may share a similar or larger fraction to neutrons
at very high densities.

In describing highly dense matter,
we clearly need information on $BB$ interaction
not only for $S=0, -1$ but also for $S \leq -2$ such as $\Xi N$,
but spectroscopic information on cascade ($\Xi$) hypernuclear systems are
severely limited at present.
%
%For $\Lambda\Lambda$ interaction, 
%the $\Lambda\Lambda$ bond energy in Lambpha (${}^6_{\Lambda\Lambda}$He)
%constrains the strength of $\Lambda\Lambda$ attraction~\cite{Tak1},
%and recent observation of enhancement in the $\Lambda\Lambda$ invariant
%mass spectrum~\cite{Yoo1} may give another aspect
%of $\Lambda\Lambda$ interaction.
%
While old emulsion data suggest a deep $\Xi^-$-nucleus potential
($\sim -24\ \mathrm{MeV}$)~\cite{Dov1},
%is about 24 MeV with Woods-Saxon type potential~\cite{Dov1}.
%that depth of
a shallow potential ($\sim -15\ \mathrm{MeV}$)
is suggested from the twin hypernuclear event
found in a nuclear emulsion~\cite{Twin}.
This shallow potential is also supported 
by the distorted wave impulse approximation (DWIA) analysis
of $\Xi^-$ production spectra in the bound state region~\cite{Fuk1,Kha1}.

In order to extract as much information as possible
from the data available at present,
we need to investingate the $\Xi^-$ hypernuclear production
spectra by $(K^-,K^+)$ reaction on nuclear targets
in both of the {\em continuum} as well as the bound state region.
Since the DWIA analysis in~\cite{Kha1} strongly rely
on the absolute value of the inclusive $\Xi^-$ production yield,
it is necessary to verify the consistency
with the production spectra in the quasi free (QF) region~\cite{Iij1}.
For this purpose,
the Green's function method of DWIA would be a useful tool,
where the continuum and bound state spectra can be described
on the same footing.

In this paper, we investigate the $\Xi^-$-nucleus potential
through $\Xi^-$ production spectra in the continuum and bound state region
in the Green's function method of DWIA
with the local optimal Fermi averaging $t$-matrix (LOFAt),
in which the $\Xi^-$-nucleus potential effects are included
in both of the strength function and the transition amplitude.
Based on the analyses of the observed continuum and bound region spectra,
we make predictions 
to the future coming $\Xi^-$ hypernuclear production experiment at J-PARC.
We find that we should observe $\Xi^-$ hypernuclear bound state peak structures
in $(K^-, K^+)$ spectra on light nuclear targets
such as $^{12}$C and $^{27}$Al 
as far as the imaginary part of the $\Xi^-$-nucleus optical potential
is not large ($|W_\Xi| \leq 3\ \mathrm{MeV}$)
and the experimental resolution is good enough
($\Delta E \leq 2\ \mathrm{MeV}$).

\section{Green's function method 
and Local optimal Fermi averaging $t$-matrix}
\label{sec:2}

The Green's function method in DWIA has been widely applied
to analyse hypernuclear reactions. 
This method has an advantage that we can describe
the continuum as well as bound state region on the same footing.
In DWIA, the differential cross section reaction
is obtained from the Fermi's golden rule~\cite{Aue1},
and the response function $R(E)$ can be decmomposed into multipole components
in the Green's function method \cite{Mor1},
\begin{eqnarray}
&&\frac{d^2\sigma}{dE_{K^+} d\Omega_{K^+}}
= \frac{p_{K^+} E_{K^+} }{(2\pi \hbar^2)^2 v_{K^-}}R(E)\ ,
\\
&&R(E)= \sum_f |\mathcal{T}_{fi}|^2\delta(E_f-E_i)\ ,
%\delta{E_{K^-}+E_T-E_{K^+}-E_H)
\nonumber\\
&&
\phantom{R(E)}=\sum_{JM\alpha\beta\alpha'\beta'}
W[\alpha\beta\alpha'\beta']
R_{\alpha\beta\alpha'\beta'}^{JM}(E)\ ,
\\
&&
R_{\alpha\beta\alpha'\beta'}^{JM}(E)=
-\frac{1}{\pi}\mbox{Im}\int r^2dr\, {r'}^2dr'
\bar{t}^*(r)
\bar{t}(r')
\nonumber\\&&
~~~~~~~~~\times
f_{JM\alpha}^*(r)
G^{JM}_{\alpha\beta\alpha'\beta'}(E;r,r')
f_{JM\alpha'}(r')
\ ,
\label{Eq:Response}
\\
&&
f_{JM\alpha}(r)= \tilde{j}_{JM}(r) \phi_{\alpha}(r)
\ ,
\\
&&W[\alpha\beta\alpha'\beta']=
(j_N\frac{1}{2}J0|j_Y\frac{1}{2})
(j'_N\frac{1}{2}J0|j'_Y\frac{1}{2})
\nonumber\\
&&~~~~\times
\delta^E_{l_N+l_Y+J}
\delta^E_{l'_N+l'_Y+J}
\sqrt{(2j_N+1)(2j'_N+1)}
%(1+(-1)^{l_N+l_Y+J})/2
%(1+(-1)^{l'_N+l'_Y+J'})/2
\ .
\end{eqnarray}
where
$v_{K^-}$ is the incident $K^-$ velocity,
subscripts $\alpha$ and $\beta$ stand for the quantum numbers of
nucleon and hyperon states, respectively,
$J$ is the total spin of hypernuclei,
$\delta^E_n=1$ and 0 for even and odd $n$,
%In the hypernuclear statistical factor $W[\alpha\beta\alpha'\beta']$,
%we define $\delta^E_n=1$ and 0 for even and odd $n$.
and $\phi_{\alpha}(r)$ is the radial wave function of the target nucleon.
Dependence on the $\Xi^-$-nucleus optical potential $U_\Xi$
appears through the Green's function $G_{\alpha\beta\alpha'\beta'}(E;r,r')$,
which contains the hypernuclear Hamiltonian.
%$H_H$.
%then we can get the information of 
%optical potential $U_Y$ between hyperon and nucleus.
%The function $\tilde{j}_{JM}$ is called distorted Bessel function \cite{Tad1}, 
The function $\tilde{j}_{JM}$ is a radial part of the product
of distorted waves $\chi^{(-)*}_{K^+}\chi^{(+)}_{K^-}$
%called distorted Bessel function \cite{Tad1}, 
evaluated in 
%with evaluating
the eikonal approximation.
We employ the $t\rho$ approximation
for the imaginary part of distortion potential,
$\mbox{Im} U_{K}(r) = \hbar v_{K} \bar{\sigma}_{KN} \rho(r)$,
where $\bar{\sigma}_{KN}$ is the isospin averaged cross sections with
$\bar{\sigma}_{NK^-}= 28.90$ mb
and $\bar{\sigma}_{NK^+}= 19.35$ mb at $P_{K^-}$=1.65 GeV/c. 
%In $(K^-,K^+)$ reactions at 1.65 GeV/$c$,
%the isospin averaged cross sections are evaluated as
%$\bar{\sigma}_{NK^-}= 28.90$ mb
%and    $\bar{\sigma}_{NK^+}= 19.35$ mb.
For the real part,
we adjust its strength to reproduce
the total cross section data of $K$ mesons~\cite{Bug1}.
%within the $t\rho$ approximation for imaginary part.
The elementary $t$-matrix elements are usually assumed to be
independent from the reaction point,
and the Fermi averaging $t$-matrix squared are factorized~\cite{Tad1}.

For $\Lambda$ and $\Sigma$ productions,
it is recently pointed out that
on-shell kinematics in the Fermi averaging (optimal Fermi averaging, OFA)
procedure roughly decide the shape of the QF spectrum~\cite{Harada}
in the Green's function method with factrized $t$-matrix,
and similar procedure for $t$-matrix was proposed in Ref.~\cite{Ale1}.
In the Semi Classical Distorted Wave (SCDW) analyses~\cite{Kohno},
the local Fermi averaging of the elementray cross section has been included.
Here we would like to incorporate both of the above two ideas;
we include the local optimal Fermi averaging $t$-matrix (LOFAt), $\bar{t}(r)$,
in the integrand of the response function Eq.~(\ref{Eq:Response}).
We define the LOFAt as,

%%%%%%%%%%%%%%%%%%%%%%%%%%%%%%%%%%%%%%%%%%%%%%%%%%%%%%%%%%%%%%%%%%%%%%%%%%%%%
%Here, we introduce {\em Local Optimal Fermi Averaging} $t$-matrix (LOFAt),
%
\begin{equation}
\bar{t} (r;\omega,\mib q)
\equiv 
\frac	{\int d\mib p_N t(s,t)\rho(p_N)
		\delta^4(P_f^\mu(r) - P_i^\mu(r))
	}
	{\int d\mib p_N \rho(p_N)
		\delta^4(P_f^\mu(r) - P_i^\mu(r))
	}\ ,
\end{equation}
where
$P^\mu_{i,f}(r)$ denote the four total momenta
in the elementary initial and final two-body states.
We adopt the Fermi distribution function
for the target nucleon momentum distribution 
$\rho(p_N)$ and parameters are taken from \cite{Aue1,All1}.
In obtaining LOFAt,
we define the $i$-th hadron single particle energy
containing the nuclear and hypernuclear potential effects as,
\begin{equation}
E_i(r)=\sqrt{\mib p_i^2+m_i^2+2m_i V_i(r)}
\sim
m_i+\frac{\mib p_i^2}{2m_i}+V_i(r)
\ .
\end{equation}
%
%\begin{equation}
%E_Y(r)=\sqrt{\boldsymbol p_Y^2+m_Y^{*2}(r)},
%\hspace{5mm}m_Y^{*2}(r)=m_Y^2+2m_Y V_Y(r)
%,
%\end{equation}
%where $i$ represents a hadron ($p$, $\Xi^-$, $K^-$ or $K^+$).
This treatment enables us to include the potential effects naturally
through the effective mass
%$m^*$, 
$m_i^{*2} = m_i^2+2 m_i V_i(r)$,
as adopted in transport models
in high-energy heavy-ion collisions~\cite{RelEng}.
Consequently,
the LOFAt has the dependence on the collision point $r$
through hadron potentials $V_i(r)$.
%It should be noted that the LOFA $t$-matrix is equivalent to ordinary 
%optimal Fermi averaging $t$-matrix when potential effects are switched off.

%In $(K^-,K^+)$ reactions at 1.65 GeV/$c$,
%the isospin averaged cross sections,
%$\bar{\sigma}_{mN}=(Z\sigma_{mp}+N\sigma_{mn})/A$, 
%are assumed to be
%$\bar{\sigma}_{NK^-}$=40mb and $\bar{\sigma}_{NK^+}$=30mb.

%%%%%%%%%%%%%%%%%%%%%%%%%%%%%%%%%%%%%%%%%%%%%%%%%%%%%%%%%%%%%%%%%%%%%%%%%%%%%%%%
\section{Results}

%%%%%%%%%%%%%%%%%%%%%%%%%%%%%%%%%%%%%%%%%%%%%%%%%%%%%%%%%%%%%%%%%%%%%%%%%%%%%%%%
% QF
%------------------------------------------------------------------------------*
\begin{figure*}
\centerline{\includegraphics[width=15.5 cm,angle=0]{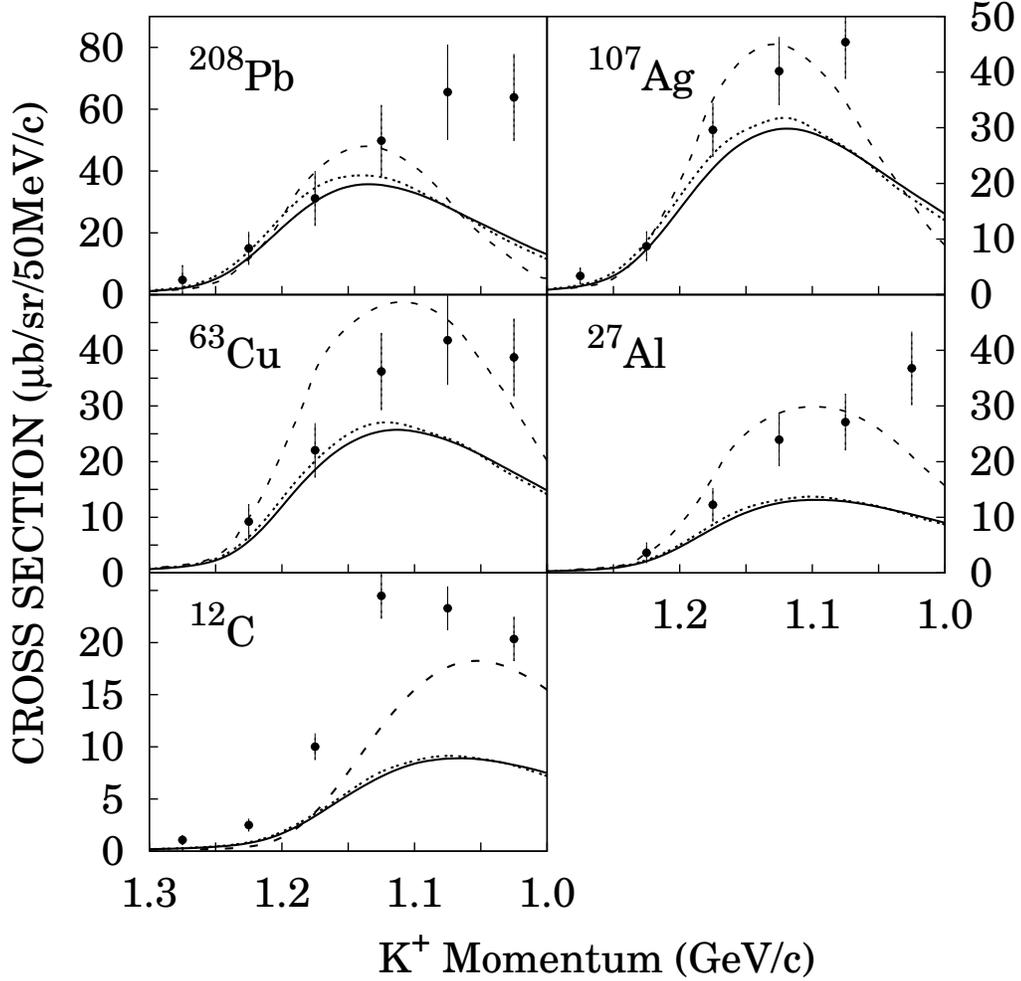}}
\caption{
Calculated $\Xi^-$-hypernuclear production spectra
in the QF region at $p_{K^-}$=1.65 GeV/$c$ and $\theta_{K^+}=6\ \mathrm{deg.}$
on C, Al, Cu, Ag and Pb targets in comparison with data~\protect{\cite{Iij1}}.
Solid lines show LOFAt + DWIA results
with $(V^\Xi_0,W^\Xi_0)=(-14 \mathrm{MeV}, -1 \mathrm{MeV})$,
and dotted lines show the results without Kaon potential effects.
In both of the calculations,
the experimental resolution is assumed
to be $\Delta E=20\ \mathrm{MeV}$ (FWHM).}
\label{Figs:Xi-QF}
\end{figure*}
%------------------------------------------------------------------------------*
In the calculation, we have assumed
the one body Woods-Saxon type hyperon-nucleus optical potential,
$U_\Xi(r) =(V_{0}^\Xi+iW_{0}^\Xi)f(r)+V_{C}^\Xi(r)$,
with 
$f(r)=1/(1+\exp((r-R)/d)$,
$R=r_0 (A-1)^{1/3}$,
$d=0.65\ \mathrm{fm}$,
$r_0=1.1\ \mathrm{fm}$,
where $V_C^\Xi(r)$ denotes $\Xi^-$-core nucleus Coulomb potential.
We assume the imaginary part of optical potential $W_0^\Xi$ to be $-1$ MeV,
which simulates the strength in the quark cluster model		%~\cite{fss2}
and the Nijmegen potential model D~\cite{Ike1} estimations.	%~\cite{fss2},
%with $d=0.65$ fm and $r_0=1.1$ fm.
%Results with a larger imaginary part of $W_0^\Xi=-3~\mathrm{MeV}$
%are also shown for comparison.
We have adopted the elementary $t$-matrix, which is re-parameterized
to fit the cross section and angle dependence for $P_{K^-}\lesssim$ 3~GeV/c.

% Xi-QF
Figure \ref{Figs:Xi-QF} shows the calculated results of $\Xi^-$ QF production spectra
with potential depth of 14 MeV
in comparison with experimental data \cite{Iij1}.
Calculated curves reproduce the experimental data systematically
on heavy targets, Cu, Ag, and Pb,
in the high $p_{K^+}$ region, where the hypernuclear excitation is small.
In the lower $p_{K^+}$ region,
other contributions have been known to be important~\cite{Gob1,Nara},
including
%$\Xi^*$ production, 
heavy-meson production and its decay,
$K^-N \to MY, M \to K^+K^-$ ($M=\phi, a_0, f_0$)~\cite{Gob1}
and 
the two-step strangeness exchange and production processes,
$K^-N \to MY, MN \to K^+Y$ ($M=\pi, \eta, \rho, ...$)~\cite{Nara}.
%charge exchange reactions,
%$K^-N \to K^0\Xi, K^0p \to K^+n$,
%and final state scattering,
%$K^-p \to K^+\Xi^-, K^+N \to K^+N$.

We underestimate the production spectra on lighter targets, $^{12}$C and $^{27}$Al.
The underestimate of QF spectrum on $^{12}$C target
is a common feature in previous DWIA calculations~\cite{Tad1,Has1}.
In Ref.~\cite{Kohno},
it is discussed that this underestimate may be due to the center-of-mass
effects:
For electron scattering on a nucleus with mass number $A$,
the center-of-mass correction in the shell model have been taken care
of by a multiplicative factor
% $F^{1/2}=\exp[q^2/(4A\alpha)]$,
$F^{1/2}=\exp[q^2/(4m_NA\hbar\omega)]$ for the form factor,
where $q$ is the momentum transfer.	% and $\alpha$ is the oscillator constant.
%With $\alpha=0.4~\mathrm{fm}^{-2}$, 
%the factor $[F^{1/2}]^2$ amounts to 1.9 and 1.3 for $^{12}$C and $^{27}$Al
%targets, respectively.
With $\hbar\omega=41 A^{-1/3} \mathrm{MeV}$,
the factor $[F^{1/2}]^2$ amounts to 1.86, 1.43, 1.22, 1.16 and 1.10
for $^{12}$C, $^{27}$Al, $^{63}$Cu, $^{109}$Ag and $^{208}$Pb targets, respectively,
at $q=500~\mathrm{MeV/c}$.
%12C:   1.85913364735732
%27Al:  1.43497056440302
%63Cu:  1.22787828785258
%107Ag: 1.15512873959884
%208Pb: 1.09700805069806
%
In Ref.~\cite{Mae1}, we have adopted a different elementary
$t$-matrix parameterization~\cite{Nara}
and larger isospin-averaged $KN$ cross sections,
then we can roughly explain the target mass dependence,
as shown with the dashed lines in Fig.~\ref{Figs:Xi-QF}.
%In the present calculation, in order to do more quantitative discussion,
%different elementary $t$-matrix is used.

%------------------------------------------------------------------------------*
\begin{figure}[tbh]
\includegraphics[width=1. \linewidth,angle=0]{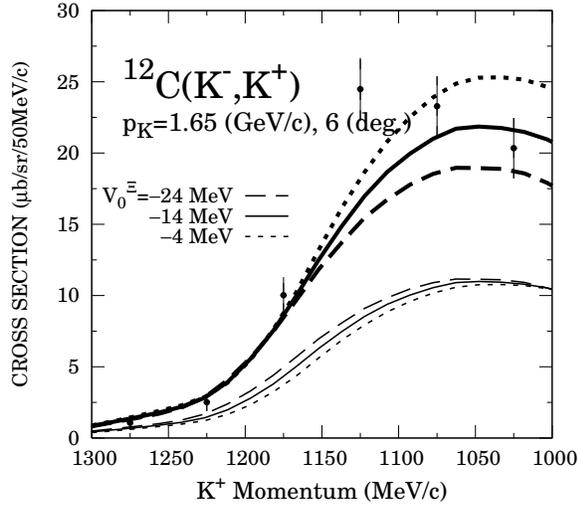} 
\caption{
Calculated $\Xi^-$ production spectra on ${}^{12}$C target
at $P_{K^-}$=1.65 (GeV/c).
%by multiplying these factors 2.35, 1.99 and 1.7 
Calculated curves are shown for the potential depth of 
$-24$ MeV (dotted), $-14$ MeV (solid) and $-4$ MeV (dashed).
Thick lines show the results with multiplicative factors
(1.7, 1.99 and 2.35 for $V_0^{\Xi} = -24, -14, -4\ \mathrm{MeV}$, respectively) 
introduced to fit the data,
and thin lines show calculated results without these factors.
%/with multiplying factors spectra (thick/thin lines).
}
\label{Fig:QF-Factor}
\end{figure}
%------------------------------------------------------------------------------*

Since our understanding is not complete
and we have several ambiguities described above
for the absolute yield in QF spectra,
we introduce an adjustable multiplicative factor
to fit the QF spectrum at low excitation energies
(high $K^+$ momentum region).
In Fig.~\ref{Fig:QF-Factor}, we show the calculated $\Xi^-$ production
spectra with $\Xi^-$ potential depth of $V_0^\Xi=-24, -14$ and $-4$ MeV
with multiplicative factors
%of 1.7, 1.99 and 2.35, respectively,
in comparison with data~\cite{Iij1}.
The potential depth dependence is small in low $p_{K^+}$ region,
and attractive potential shifts the spectrum
towards the high $p_{K^+}$ direction slightly.
When we adjust the multiplication factors as described above,
calculated results reasonably well explain the QF spectrum.
This means that we cannot determine the potential depth
accurately from the QF spectrum shape.

%------------------------------------------------------------------------------*
\begin{figure}[htbp]
\begin{center}
\includegraphics[width=1. \linewidth,angle=0]{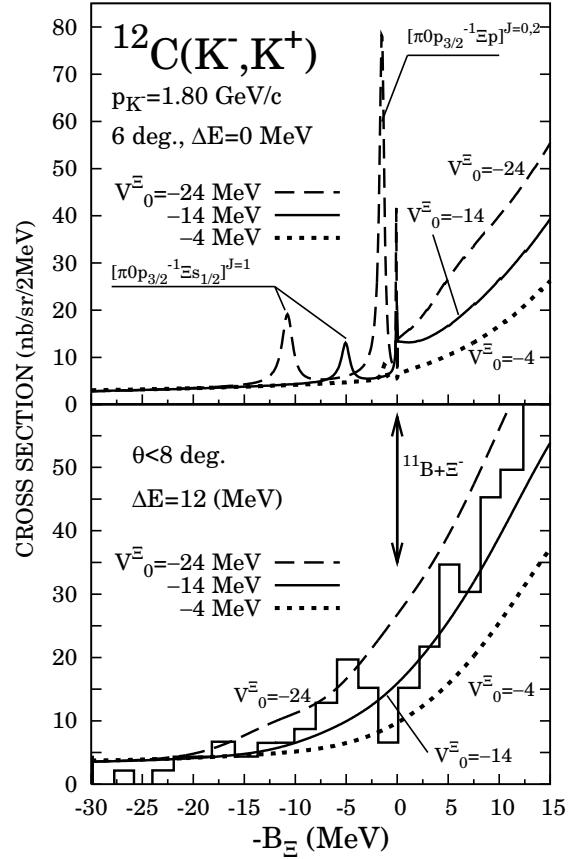} 
\end{center}
\caption{
Potential depth dependence of the $\Xi^-$-hypernuclear production spectra
in the bound state region at $p_\pi$=1.80 GeV/$c$ and $\theta\leq$ 8 deg.
on ${}^{12}$C without(with) the experimental resolution (upper/lower panel).
Dotted, solid and dashed lines show the results with
$\Xi^-$-nucleus potential depths of 24, 14 and 4 MeV, respectively.
%In the lower panel, 
Experimental data are taken from Ref.~\protect{\cite{Kha1}}.
}
\label{FigVdep}
\end{figure}
%------------------------------------------------------------------------------*
%------------------------------------------------------------------------------*
\begin{figure*}[tbh]
\centerline{
\includegraphics[width=0.4\linewidth,angle=-90]{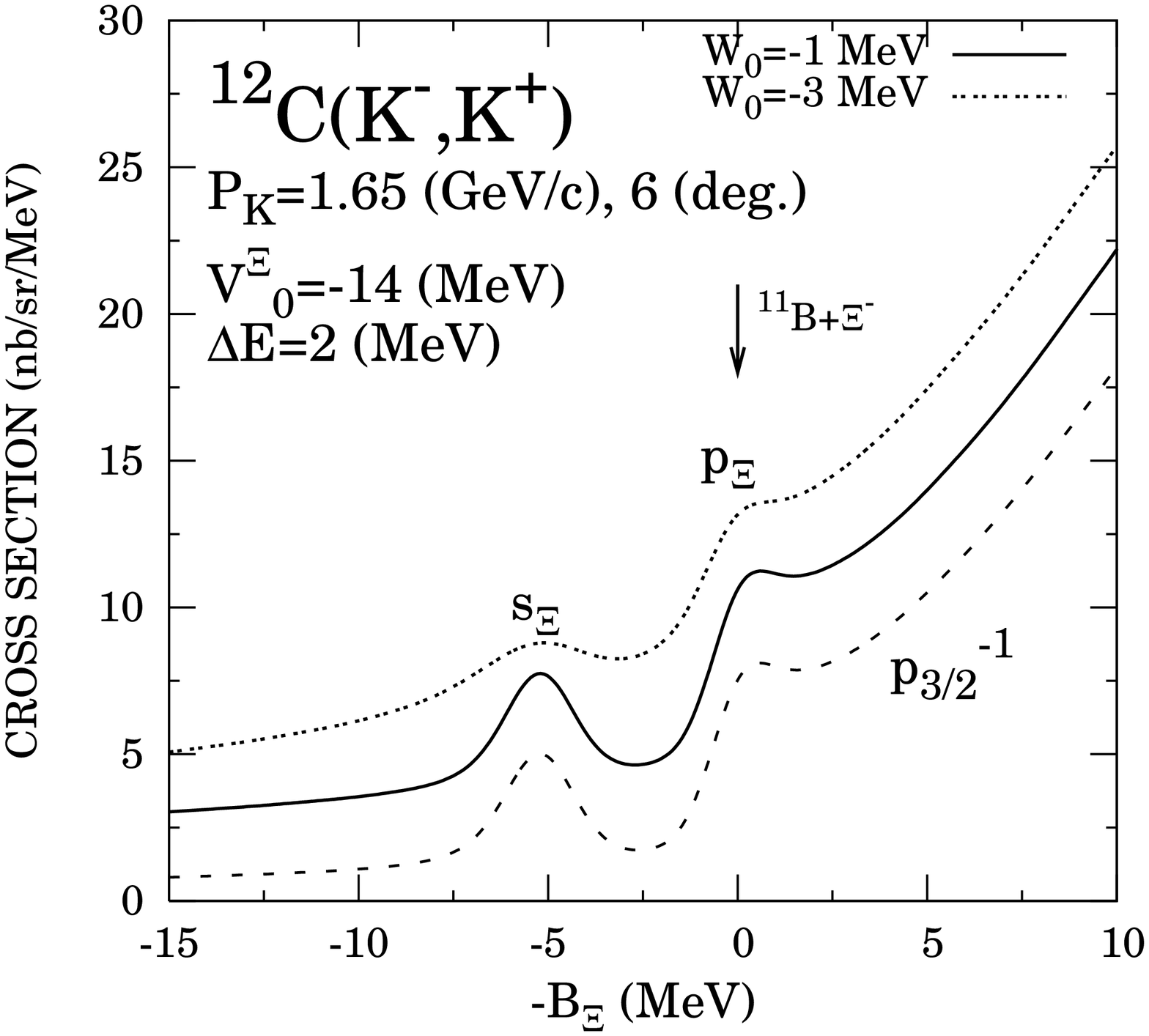}%
\includegraphics[width=0.4\linewidth,angle=-90]{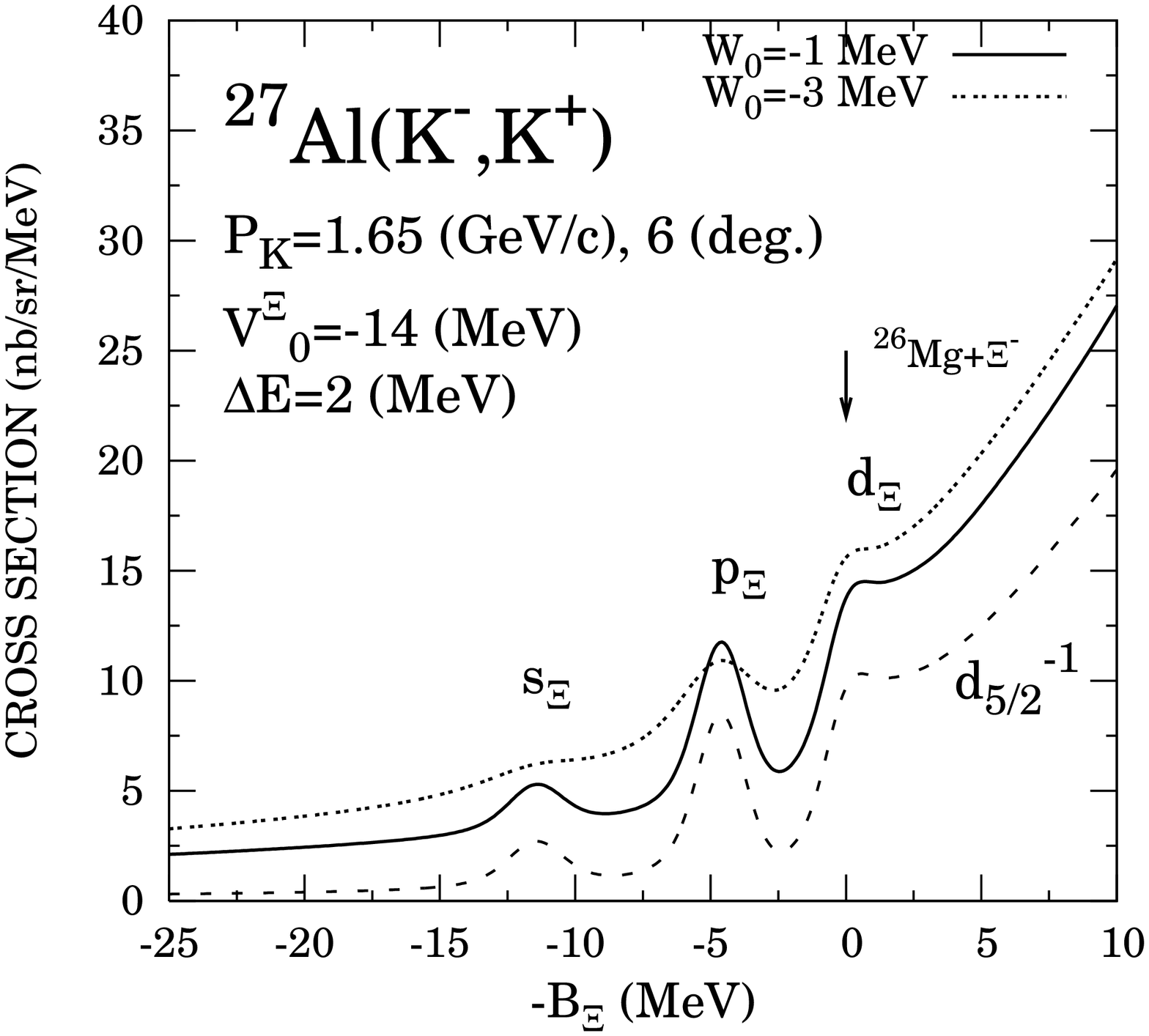} 
}
\centerline{
\includegraphics[width=0.4\linewidth,angle=-90]{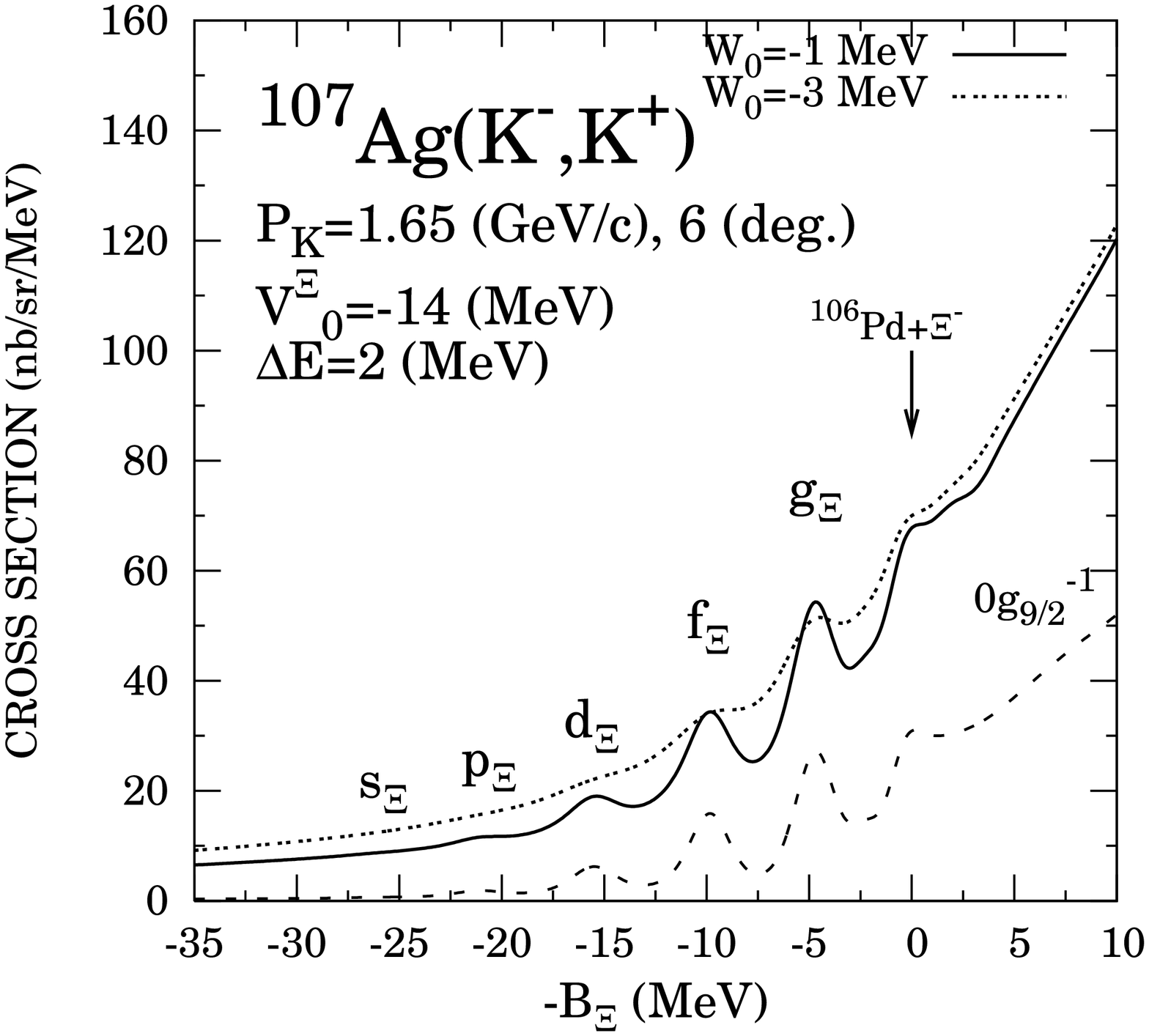}%
\includegraphics[width=0.4\linewidth,angle=-90]{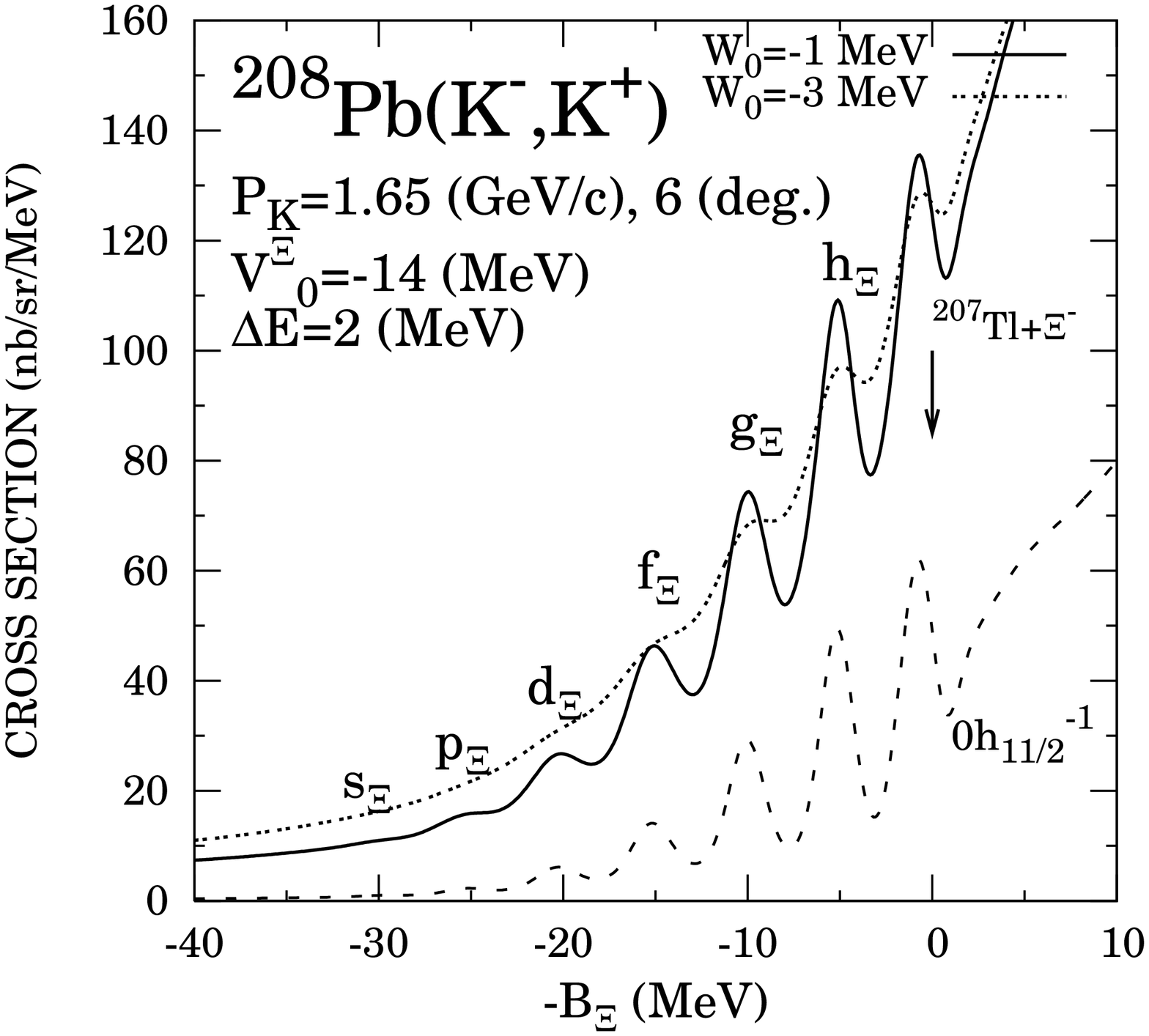} 
}
\caption{
$\Xi^-$ production spectra
at $p_{K^-}=1.65\ \mathrm{GeV}/c$ and $\theta_{K^+}$=6 deg.
on $^{12}$C, $^{27}$Al, $^{107}$Ag and $^{208}$Pb targets
expected in the J-PARC experiment.
We assume a Woods-Saxon potential with 14 MeV depth,
and we show the results with imaginary parts
of $-1$ MeV (solid) and $-3$ MeV (dotted).
Experimental resolution is assumed to be $\Delta E=2\ \mathrm{MeV}$ (FWHM).
}
\label{BS}
\end{figure*}
%------------------------------------------------------------------------------*

On the other hand, production spectrum is more sensitive to the potential
depth at low excitation energies,
then there is a possibility that we can determine the potential depth
%from the $\Xi^-$ production spectra in the bound state region,
as discussed in Refs.~\cite{Fuk1}.
In Fig.~\ref{FigVdep},
we show the results of the potential depth dependence
of the calculated $\Xi^-$ production spectrum on ${}^{12}$C target
at $p_{K^-}=1.80$ GeV/$c$
multiplied by the factors described above
in comparison with data~\cite{Fuk1}.

Ikeda {\em et al.} evaluated
the width of $\Xi^-$ hypernuclear states (${}^{11}$B+$\Xi^-$)
to possible double $\Lambda$ states
based on the Nijmegen model D potential as 1.2 MeV and 0.5 MeV
for $[\pi 0p_{3/2}^{-1}\otimes \Xi s_{1/2}]^{J=1}$ and
$[\pi 0p_{3/2}^{-1}\otimes \Xi p]^{J=0,2}$,
respectively~\cite{Ike1}.
In the upper panel of Fig.~\ref{FigVdep},
we show the ideal $\Xi^-$ production spectra on ${}^{12}$C target
without the energy resolution folding.
The imaginary part is assumed to be $W^\Xi_0=-1$~MeV.
These calculated spectra show that the $\Xi^-$ hypernuclear state widths are
in good agreements with the estimates in Ref.~\cite{Ike1}.
In comparison with experimental data,
these spectra must be folded using a Gauss function. 
%%%%%%%%%%%%%%%%%%%%%%%%%%%%%%%%%%%%%%%%%%%%%%%%%%%%
In the lower panel of Fig.~\ref{FigVdep},
we show the results with an experimental resolution of
$\Delta E=12\ \mathrm{MeV}$ FWHM.
Since the experimental resolution is not enough
to distinguish the bound state peaks and the statistics is low,
we should compare the integrated yield in the bound state region.
We find clear potential dependence in the bound state region,
and with deep $\Xi^-$ potential ($U_\Xi=-24~\mathrm{MeV}$) we may find
a bump structure at around $-B_\Xi \sim -12 \mathrm {MeV}$
even with this low resolution.
Comparison with the data suggests that 
the potential depth around 14 MeV is preferred in the present treatment,
and if the statistics is high enough,
it would be possible to determine the potential depth even with low resolution.

%%%%%%%%%%%%%%%%%%%%%%%%%%%%%%%%%%%%%%%%%%%%%%%%%%%%%%%%%%%%%%%%%%%%%%%%%%%%%%%%
% Xi-bound state region
The depth of the $\Xi^-$-nucleus potential has been already suggested to be
around 14 MeV from the analysis of the $(K^-,K^+)$ spectrum
in the bound state region~\cite{Fuk1}.
In that analysis,
the $t$-matrix element is evaluated
under the frozen nucleon momentum approximation,
where the kinematics is given with zero initial nucleon momentum.
In the present analysis,
while the kinematics in the elementary process
and the way to fix the absolute value
are different %from those in Ref.~\cite{Fuk1},
preferred potential depth is similar.
This may be due to the fact that
the excitation energy dependence of LOFAt is weak and smooth
since the covered $K^+$ momentum range is narrow in the bound state region.

%%%%%%%%%%%%%%%%%%%%%%%%%%%%%%%%%%%%%%%%%%%%%%%%%%%%%%%%%%%%%%%%%%%%%%%%%%%%%%%%
% Xi-bound state peaks
%%%%%%%%%%%%%%%%%%%%%%%%%%%%%%%%%%%%%%%%%%%%%%%%%%%%%%%%%%%%%%%%%%%%%%%%%%%%%%%%
Now we find that $\Xi^-$-nucleus potential with 14 MeV depth
well describes the spectra in the bound state and QF region
for light nuclear targets,
then it would be valuable to predict the peak structure
which would be observed in the future coming J-PARC day-one experiment.
Sensitivity of calculated spectra for the $\Xi^-$-nucleus potential 
is weak in the QF region
because of the high momentum transfer $q\sim$ 500 MeV/c,
therefore it is difficult to extract precise potential information
from the QF region, as shown in Fig.~\ref{Fig:QF-Factor}

%%%%%%%%%%%%%%%%%%%%%%%%%%%%%%%%%%%%%%%%%%%%%%%%%%%%%%%%%%%%%%%%%%%%%%%%%%%%%%%%

%\subsection{Prediction spectra at J-PARC}

In Fig. \ref{BS},
we show the calculated $K^+$ spectra in the bound state region
of $(K^-,K^+)$ reactions
on ${}^{12}$C, ${}^{27}$Al, ${}^{107}$Ag and ${}^{208}$Pb targets
with a potential depth of $V^\Xi_0=-14\ \mathrm{MeV}$,
which explains the QF spectra and low resolution spectra in the bound region.
We compare the results with
$W^\Xi_0=-1~\mathrm{MeV}$ (solid lines)
and 
$W^\Xi_0=-3~\mathrm{MeV}$ (dotted lines).
%Hole state contribution (dashed line)
We assume that the experimental resolution of $\Delta E=2\ \mathrm{MeV}$
would be achieved.
%Solid line shows LOFAt + DWIA results using $\Xi^-$-nucleus potential depth of 15 MeV 
%and imaginary part of 1 MeV.
%10 MeV in Fig. \ref{Figs:Xi-QF} and 2 MeV in Fig. \ref{Fig:Xi-BS}.
%
We find that bound state peaks are populated selectively
due to high momentum transfer $(q\sim 500~\mbox{MeV/c})$
as in the $\Lambda$ production spectra by $(\pi^+,K^+)$ reactions
($q\sim 350\ \mathrm{MeV}/c$),
and these peaks can be identified
in the high resolution experiment.% with $\Delta E=2\ \mathrm{MeV}$.
%if the experimental resolution is
%improved to be around 2 MeV.

In the Green's function method,
target nucleon deep hole states
are assumed to have large imaginary energies.
Therefore,
calculated results may be overestimating
the spectra around the ground state
%$E^*\sim 0$~MeV
due to the long Lorentzian tail from the deep hole states
having finite contributions in this energy region.
This problem will be discussed in the future.

\section{Conclusion}

We have investigated the $\Xi^-$-nucleus potential
through cascade ($\Xi$) hypernuclear production spectra by $(K^-,K^+)$ reaction
in the Green's function method~\cite{Mor1}
of the distorted wave impulse approximation (DWIA)
with the local optimal Fermi averaging $t$-matrix (LOFAt)~\cite{Mae1}.
The calculated spectra are in good agreement 
with the experimental data for heavy targets.
With the multiplicative factor adjusted to fit the spectra
on light targets,
we find that the calculated spectra
well reproduce the observed spectrum~\cite{Kha1} in shape and yield
with $\Xi^-$-nucleus potential depth around 14 MeV.
This potential depth is consistent with those suggested
in previous works~\cite{Twin,Fuk1,Kha1}.

While the dependence on the potential depth is small in the continuum region,
it is clealy distinguished in the bound region.
Therefore, it would be possible to extract the $\Xi^-$-nucleus potential depth
from the production yield in the bound state region
when the statistics is high enough.
Furthermore, the $\Xi^-$ bound state peak structure can be found
in the $(K^-,K^+)$ spectra on light target such as $^{12}$C and $^{27}$Al,
as far as the imaginary part is not very large
($|W_0^\Xi| \leq 3\ \mathrm{MeV}$)
and the experimental resolution is improved ($\Delta E \sim 2\ \mathrm{MeV}$),
as expected in the J-PARC experiment.
We believe that our prediction would provide useful information
in searching for the $\Xi^-$ nuclear bound states at J-PARC.

\section*{Acknowledgements}

We would like to thank Prof. A. Gal, Prof. T. Harada and Prof. M. Kohno
for valuable discussions.
This work is supported in part by the Ministry of Education,
Science, Sports and Culture,
Grant-in-Aid for Scientific Research under the grant numbers,
15540243,	% Kiban(C)(Ohnishi,2003-2006)
1707005,	% Tokutei
and 19540252.	% Kiban(C)(Ohnishi,2007-2009)

%as required. Don't forget to give each section
%and subsection a unique label (see Sect.~\ref{sec:1}).
%

%
% BibTeX users please use
% \bibliographystyle{}
% \bibliography{}
%\begin{thebibliography}{}

\end{document}